\documentclass{aa}
\usepackage{psfig}

\def\ltsima{$\; \buildrel < \over \sim \;$}
\def\lsim{\lower.5ex\hbox{\ltsima}}
\def\gtsima{$\; \buildrel > \over \sim \;$}
\def\gsim{\lower.5ex\hbox{\gtsima}}

\begin{document}

\title{The {\it BeppoSAX} view of the galactic \\
high-mass X--ray binary 4U 0114+65}

\author{N. Masetti\inst{1} 
\and
M. Orlandini\inst{1}
\and
D. Dal Fiume\inst{2,}\thanks{Deceased.}
\and
S. Del Sordo\inst{3}
\and \\
L. Amati\inst{1}
\and
F. Frontera\inst{1,4}
\and
E. Palazzi\inst{1}
\and
A. Santangelo\inst{5}
}

\institute{
Istituto di Astrofisica Spaziale e Fisica Cosmica di Bologna, 
INAF, via Gobetti 101, I-40129 Bologna, Italy (formerly IASF/CNR, 
Bologna)
\and
Istituto Tecnologie e Studio sulla Radiazione Extraterrestre, CNR,
via Gobetti 101, I-40129 Bologna, Italy
\and
Istituto di Astrofisica Spaziale e Fisica Cosmica di Palermo, 
INAF, via La Malfa 153, I-90146 Palermo, Italy (formerly IASF/CNR, Palermo)
\and
Dipartimento di Fisica, Universit\`a di Ferrara, via Paradiso 12, I-44100
Ferrara, Italy
\and
Institut f\"ur Astronomie und Astrophysik, Eberhard Karls Universit\"at 
T\"ubingen, Sand 1, D-72076 T\"ubingen, Germany
}

\titlerunning{A {\it BeppoSAX} observation of the HMXB 4U 0114+65}
\authorrunning{Masetti et al.}

\offprints{N. Masetti, {\tt masetti@bo.iasf.cnr.it}}

\date{Received 17 June 2005; Accepted 6 September 2005}

\abstract{A pointed observation of the galactic high-mass X--ray binary
4U 0114+65 was carried out with {\it BeppoSAX} to compare the
X--ray spectral and timing characteristics observed by this satellite
over the broadest range of energies thus far (1.5--100 keV) with the
information previously obtained with other spacecraft. The light curve
of 4U 0114+65 shows a large flare at the beginning of the {\it
BeppoSAX} pointing and no significant hardness evolution either during
the flare or in the low state occurring after the flare itself. The
modulation at $\sim$2.7 hours, attributed to the accreting neutron star
(NS) spin periodicity, is not significantly detected in our data,
although fluctuations with timescales of $\sim$3 hours can be seen in
the 2--10 keV light curve. Shorter modulations down to timescales of
minutes are also found and interpreted as due to accretion of matter
onto the NS. The flaring and the low state spectra of 4U 0114+65 can be
equally well fitted either with a power law modulated by a high-energy
exponential cutoff or with a Comptonization model. During the low state
the presence, although tentative, of a thermal component (with $kT
\sim$ 0.3 keV) at low energies, possibly produced by an ionized plasma
cloud around the NS, cannot be excluded. Contrary to previous claims, a
cyclotron resonant feature in absorption at $\sim$22 keV was not
detected in the {\it BeppoSAX} spectroscopic data, whereas evidence for a
Fe emission line around 6.4 keV is found only during the low state
emission. Using all of the above information, a scenario for the system
in which the NS is embedded in, and accreting from, a low angular
momentum gas cloud is envisaged.
}

\maketitle

\section{Introduction}

Persistently emitting High-Mass X--ray Binaries (HMXBs) are
gravitationally bound systems composed of a early-type star losing matter,
generally via stellar wind, in favour of a compact object, a neutron star
(NS) or possibly a black hole. This accretion phenomenon induces
high-energy emission from the accreting flow and, in the case of a NS as
the accretor, from the surface of the compact object. This kind of HMXB
has X--ray luminosities of the order of at least 10$^{35}$ erg s$^{-1}$ in
the 2--10 keV range. Also, if the accreting object is a magnetized NS,
the accretion flow is channeled onto the magnetic polar caps and the
X--ray emission can appear as pulsed, modulated by the NS spin period (see
White et al. 1995 for a review).

The hard X--ray source 4U 0114+65 belongs to this group of X--ray
binaries. It was discovered by the {\it SAS-3} satellite (Dower et al.  
1978) and extensively observed in the X--ray band by several spacecraft.

The lightcurve of the object appears quite variable in X--rays, showing 
flares lasting a few hours, occurring on daily timescales and that 
are 15-20 times brighter than the persistent low-level emission (Yamauchi 
et al. 1990; Apparao et al. 1991). Shorter-term flickering on timescales 
of minutes was also observed (Koenigsberger et al. 1983). Finley et al. 
(1992), by analyzing archival {\it EXOSAT} and {\it ROSAT} data, showed 
the existence of a very stable 2.78-hour periodicity in the X--ray light 
curve of 4U 0114+65; this was more recently confirmed by Corbet et al. 
(1999) and Hall et al. (2000) by using {\it RXTE} ASM and PCA data, 
respectively. This was interpreted by these authors in terms of the the 
spin period of a slowly rotating NS. Farrell et al. (2005) moreover 
reported the presence of a superorbital periodicity of 30.7 days in this 
system, again by the analysis of {\it RXTE}/ASM data. This may 
suggest the presence of a precessing accretion disk around the NS.

X--ray spectroscopy (Yamauchi et al. 1990; Apparao et al. 1991; Hall et
al. 2000) shows that the source spectrum is satisfactorily fitted with
the `classical' phenomenological model for accreting NSs in HMXBs 
(White et al. 1983), i.e. a power law modified at high energies by an
exponential cutoff. In the case of 4U 0114+65, typical spectral
parameter values are found to be $\Gamma \sim$ 1, $E_{\rm cut} \sim$ 8
keV and $E_{\rm fold} \sim$ 20 keV for the photon index, the cutoff and
folding energies. A neutral iron emission line at 6.4 keV was also 
sometimes detected, preferentially during the persistent low state 
emission.

Bonning \& Falanga (2005) observed this source with {\it INTEGRAL} and 
again found the 2.7-hour spin modulation up to an energy of
80 keV. The 5--100 keV {\it INTEGRAL} spectrum was consistent with the
findings, illustrated above, of previous high-energy observations. They
also measured a NS spin-up trend $\dot{P}$ = $-$8.9$\times$10$^{-7}$ s
s$^{-1}$ over $\sim$8 years. Moreover, these authors tentatively 
identified a Cyclotron Resonant Feature (CRF) at $\sim$22 keV in the 
X--ray spectrum of the source, implying a magnetic field of
2.5$\times$10$^{12}$ G for the accreting NS.

This evidence can be interpreted as due to accretion onto a NS from a 
wind coming from an early-type star. Indeed, 4U 0114+65 was
associated with its optical counterpart, the 11$^{\rm th}$ magnitude 
star V662 Cas (also known as LSI +65$^{\circ}$010) by Margon \& Bradt 
(1977).
Subsequently, Crampton et al. (1985) spectroscopically determined the
orbital period of the system as $P_{\rm orb}$ = 11$\fd$59; this modulation
was also found later in the {\it RXTE}/ASM X--ray data by Corbet et al.
(1999). Reig et al. (1996) performed a thorough study of the optical
properties of the companion star and found it is a blue supergiant of
spectral type B1Ia located at 7.0$\pm$3.6 kpc from earth, though 
Koenigsberger et al. (2003) note that it could be located closer. 
This star is in a basically circular orbit (Crampton et al. 1985; 
Koenigsberger et al. 2003) and may induce X--ray eclipses 
(Hall et al. 2000).

The above optical information conclusively placed 4U 0114+65 among
wind-accreting supergiant HMXBs in terms of spectral type of the 
companion
and of X--ray luminosity budget, the latter being between 10$^{35}$ and
10$^{36}$ erg s$^{-1}$ in the 2--10 keV range; indeed, previous distance
estimates ($d \sim$ 2 kpc; e.g., Aab et al. 1983) led to a 2--10 keV
luminosity of about ten times lower, erroneously implying that this source
was a low-luminosity HMXB.

No radio emission was detected from this source in the 5 GHz band down to 
a level of 1.2 mJy (Nelson \& Spencer 1988).

We observed 4U 0114+65 with {\it BeppoSAX} (Boella et al. 1997a) within
the framework of our program of pointed observations of HMXBs (Dal 
Fiume
et al. 2000; Orlandini et al. 1998,1999,2000,2004; Masetti et al. 2004) in
order to describe its X--ray temporal and spectral behaviour over the wide
range of energies (0.1--300 keV) covered by this satellite, and
to compare its X--ray characteristics with those of other objects
belonging to this class.

Here we present the results of a pointed observation made with {\it
BeppoSAX} on this system in January 1999. Thanks to the
broadband spectroscopic capabilities of this satellite and to its high
sensitivity, for the first time we were able to simultaneously explore the
1.5--100 keV emission from this source. The paper is organized as follows:
Sect. 2 describes the observations and the data analysis, in Sect. 3 the 
results showing the X--ray spectral and timing behaviour of 4U 0114+65 
will be reported; in Sect. 4 a discussion is given.

\section{The {\it BeppoSAX} observation}

4U 0114+65 was observed between January 26 and 27, 1999, for $\sim$63 ks, 
by the four coaligned Narrow-Field Instruments
(NFIs) carried by {\it BeppoSAX}: the Low Energy Concentrator Spectrometer
(LECS, 0.1--10 keV; Parmar et al. 1997), two Medium Energy Concentrator
Spectrometers (MECS, 1.5--10 keV; Boella et al. 1997), the High-Pressure
Gas Scintillation Proportional Counter (HPGSPC, 6--60 keV; Manzo et al.
1997) and the Phoswich Detection System (PDS, 15--300 keV; Frontera et al.
1997). The total duration of this {\it BeppoSAX} pointing along with the
on-source exposure times for each NFI used are reported in Table~1.

\begin{table*}
\caption[]{Log of the {\it BeppoSAX} observation presented in this paper.}
\begin{center}
\begin{tabular}{lcccccc}
\noalign{\smallskip}
\hline
\noalign{\smallskip}
\multicolumn{1}{c}{Start day} & Start time & Duration &
\multicolumn{4}{c}{On-source time (ks)} \\
 & (UT) & (ks) & LECS & MECS & HPGSPC & PDS \\
\noalign{\smallskip}
\hline
\noalign{\smallskip}
 1999 Jan 26 & 21:38:42 & 63.2 &  8.4 & 31.1 & 33.9 & 16.2 \\
\noalign{\smallskip}
\hline
\noalign{\smallskip}
\end{tabular}
\end{center}
\end{table*}

\begin{figure*}[th!]
\hspace{-0.2cm}
\psfig{figure=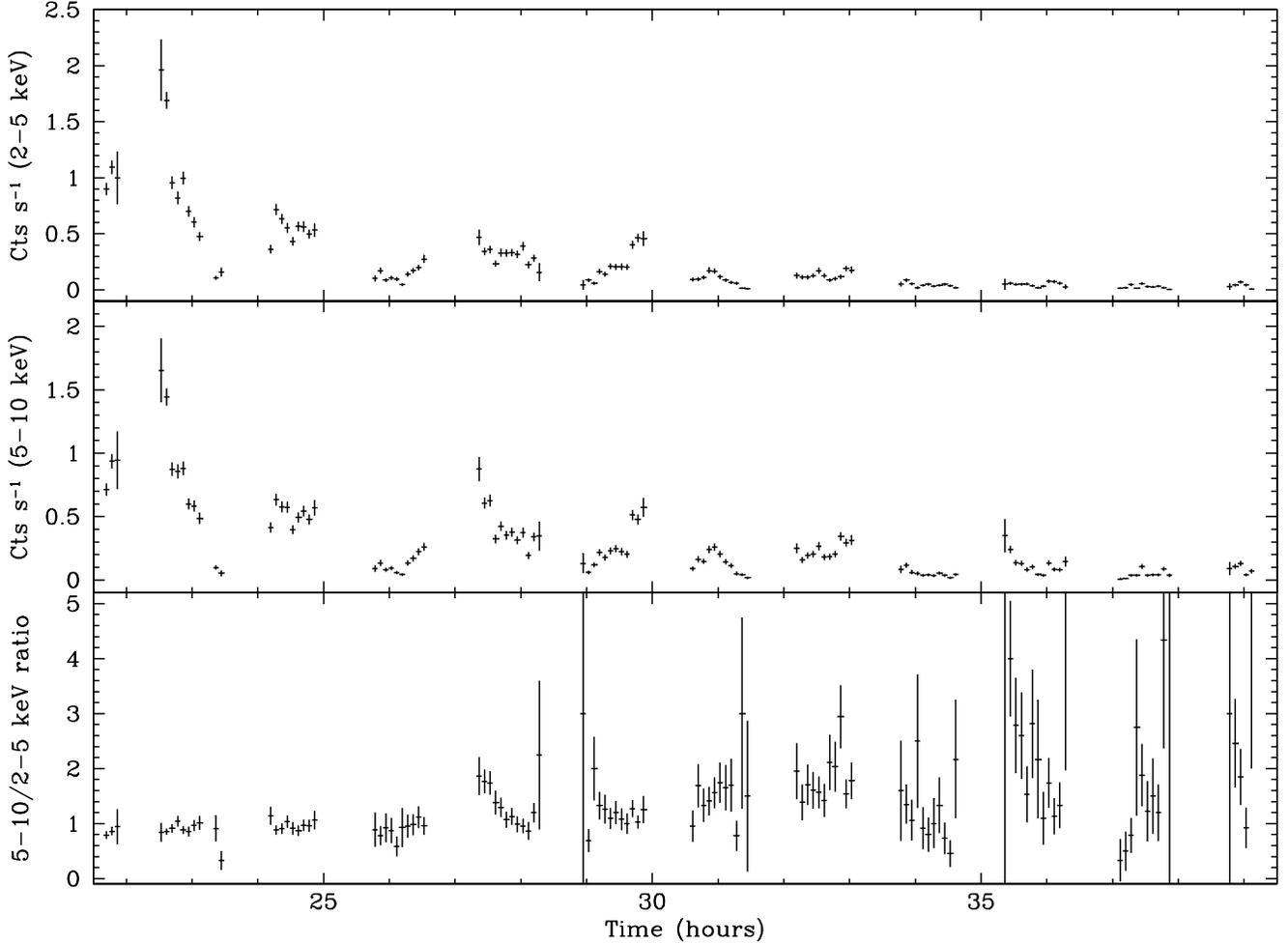,width=18.4cm,angle=-90}
\vspace{-0.7cm}
\caption[]{Background-subtracted X--ray light curves of 4U 0114+65 in the
2--5 keV {\it (upper panel)}, 5--10 keV {\it (middle panel)} bands, along
with their ratio {\it (lower panel)} as observed during the {\it BeppoSAX}
pointing. All curves are rebinned at 300 s. A large flare during the first
3 hours of the observation is apparent. Subsequent flickering activity
down to timescales of minutes can be noticed. Time is measured from 
the start of the {\it BeppoSAX} observation, on January 26, 1999, 
at 21:38:42 UT (see also Table 1), and it is expressed in hours
starting from 00:00 UT of January 26, 1999. Vertical error bars show
1$\sigma$ confidence level uncertainties for each bin.}
\end{figure*}

The source was well detected by all four NFIs.
Good NFI data were selected from intervals outside the South Atlantic
Geomagnetic Anomaly when the elevation angle above the earth limb was
$>$$5^{\circ}$, when the instrument functioning was nominal and, for LECS
events, during spacecraft night time. The SAXDAS 2.0.0 data analysis
package (Lammers 1997) was used for the extraction and the processing of
LECS and MECS data. The PDS data reduction was performed
using XAS version 2.1 (Chiappetti \& Dal Fiume 1997).
LECS and MECS data were both reduced using an extraction radius of 4$'$
centered on the source position; before extraction, data from the two MECS
units were merged. Background subtraction for the two imaging instruments
was performed using standard library files. The background for the 
PDS data was evaluated from the fields observed during off-source pointing
intervals, whereas that of the HPGSPC was computed through an
Earth-occultation technique (Manzo et al. 1997).

Because 4U 0114+65 is located near the Galactic plane and is not a
particularly bright X--ray binary, we checked for possible effects induced
by Galactic diffuse emission in the PDS data background evaluation. The
PDS off-source fields for background evaluation were indeed at different
Galactic latitudes ($b$ = $-$0$\fdg$9 and $b$ = +6$\fdg$0) with
respect to the source ($b$ = +2$\fdg$6), so a possible gradient in 
the Galactic diffuse emission, if present, can be measured and 
eventually removed from the PDS data. However, the count rate
difference between the two fields is $-$0.06$\pm$0.07 counts s$^{-1}$, 
thus consistent with zero; moreover, this difference impacts on the 
background estimate by less than 0.3\%. Therefore we considered this 
effect negligible.

\section{Results}

\subsection{Light curves}

We extracted the 2--5 and 5--10 keV light curves of 4U 0114+65 from the
{\it BeppoSAX}/MECS data. These background-subtracted MECS light curves
rebinned at 300 s, along with their ratio, are shown in Fig. 1.  A
large flare, similar to those reported by Apparao et al. 
(1991), was seen from 4U 0114+65 at the beginning of the {\it
BeppoSAX} observation (Fig. 1). Its peak intensity is $\sim$20 times the
persistent low-level emission observed after the event. No significant
hardness evolution was detected during the episode, although the (5--10
keV)/(2--5 keV) hardness ratio seems to increase starting $\sim$5 hours
after the beginning of the observation, just after the end of the flare
(Fig. 1, lower panel). The light curves also show substantial random
variability in the form of internal fluctuations lasting from a few 
hours to minutes (the so-called `X--ray flickering').

In order to see whether this erratic variability implied spectral changes
depending on the source intensity, we computed a hardness-intensity ratio
between the 5--10 keV and the 2--5 keV count rates and plotted it against
the total 2--10 keV count rate. The results are plotted in Fig. 2:
although the figure hints to an inverse correlation between the two 
quantities, there is no significant dependence of the hardness 
ratio on the total intensity of 4U 0114+65 in the 2--10 keV range, as 
opposed to the inverse relation between X--ray hardness and intensity 
found by Apparao et al. (1991) from {\it EXOSAT} data.

Using the optical orbital ephemeris reported by Crampton et al. 
(1985) it is found that this large flare occurred just after periastron 
passage: in detail, we find that the flare occurred at phase $\phi \sim$ 
0.05 assuming the most likely case of circular orbits for the system 
(Koenigsberger et al. 2003).

\begin{figure}
\psfig{figure=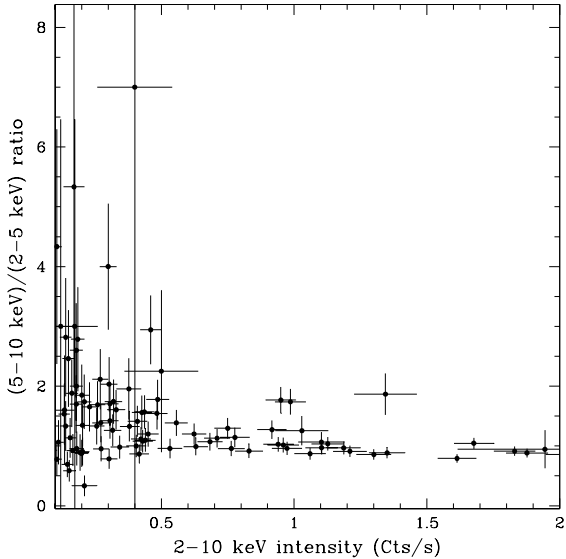,width=8.8cm,angle=0}
\vspace{-0.7cm}
\caption[]{Hardness-intensity diagram for the 2--10 keV emission of 4U
0114+65 during the {\it BeppoSAX} pointing. Although an hint of an 
inverse correlation between the 5--10 keV / 2--5 keV hardness ratio and
the 2--10 keV intensity is suggested by the plot, no statistically 
significant trend is present. Error bars show 1$\sigma$ confidence level 
uncertainties for each data point.}
\end{figure}

\subsection{Timing properties}

Timing analysis on the 2--10 keV data (where the S/N ratio was highest)
was performed with the FTOOLS v5.1\footnote{available at: \\ {\tt
http://heasarc.gsfc.nasa.gov/ftools/}} (Blackburn 1995)
tasks {\tt powspec} and {\tt efold}, after having converted the event
arrival times to the solar system barycentric frame.

The results do not reveal the presence of any kind of periodicity or
quasiperiodic oscillations. We were not able to detect the 2.7 hour NS
spin periodicity in our analysis. The Power Spectral Density (PSD; Fig. 3)
obtained with this analysis is characterized by red noise and shows no
significant deviations from the 1/$f$-type distribution (where $f$ is the
time frequency) typical of the `shot-noise' behaviour. This trend however
flattens for frequencies lower than $f \approx$ 10$^{-4}$ Hz, which is
of the order of magnitude of the 4U 0114+65 NS spin frequency. Indeed, one
can see from Fig. 1 that, at least in the first half of the {\it BeppoSAX}
observation, the presence of fluctuations on a timescale of $\sim$3 hours
is suggested. We stress that the non-detection of this modulation may 
be due to the relatively short duration of the observation ($\sim$3 pulse 
cycles if one considers the on-source time), to the sparse light curve 
sampling and to the presence of the large flare at the beginning of the 
pointing.

\begin{figure}
\hspace{-.5cm}
\psfig{figure=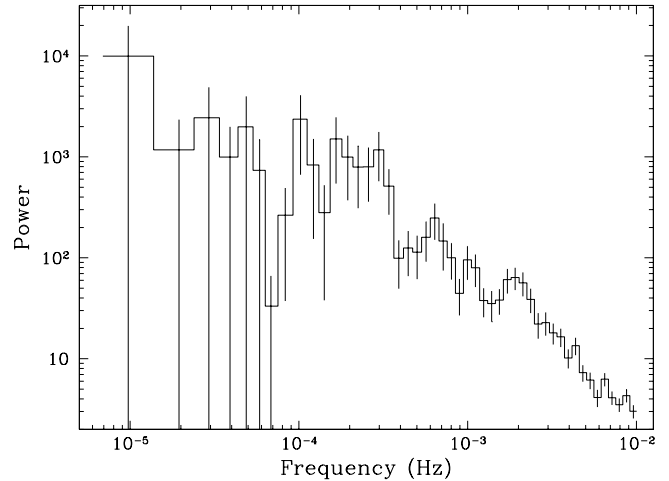,width=10cm,angle=-90}
\vspace{-0.7cm}
\caption[]{2--10 keV PSD of 4U 0114+65 obtained from the events 
registered by the two MECS units. The PSD is normalized according to the 
prescription by Leahy et al. (1983). No coherent oscillation or 
periodicity is found, whereas a shot-noise 1/$f$-type trend is apparent, 
flattening at frequencies below $f \approx$ 10$^{-4}$ Hz.}
\end{figure}

\subsection{Spectra}

\begin{table}
\caption[]{Best-fit spectral parameters for the flaring and low-level
emission of 4U 0114+65. In the leftmost column of the table,
$N_{\rm H}$ is the neutral hydrogen column density, $\Gamma$ is the power 
law photon index, $E_{\rm cut}$ is the cutoff energy, $E_{\rm fold}$
is the exponential folding energy, $T_{\rm 0}$ is the temperature of the 
Comptonized seed photons, $T_{\rm e^-}$ is the temperature of the 
Comptonizing electron plasma, $\tau$ is the optical depth of the 
Comptonizing plasma, and the $K$s are the normalization constants of each 
model. Quoted errors are at the 90\% confidence level 
for a single parameter. Luminosities, corrected for interstellar Galactic
absorption, are computed assuming a distance $d$ = 7.0 kpc and using the 
{\sc HighECutPL} model as reference; no significant differences are 
found if other models are used.}
\begin{center}
\begin{tabular}{l|r|r}
\hline
\hline
\noalign{\smallskip}
Model + parameters & \multicolumn{1}{c|}{Flare} & 
\multicolumn{1}{c}{Low-level} \\
\noalign{\smallskip}
\hline
\noalign{\smallskip}

{\sc CutoffPL}: & & \\
$N_{\rm H}$$^{a}$ & 8.9$^{+0.8}_{-0.7}$ & 17.6$^{+1.6}_{-1.7}$ \\
$\Gamma$ & 0.86$\pm$0.15 & 0.92$^{+0.19}_{-0.23}$ \\
$E_{\rm cut}$ (keV) & 15$^{+3}_{-2}$ & 18$^{+5}_{-4}$ \\
$K_{\rm PL}$ ($\times$10$^{-2})$ & 2.2$^{+0.6}_{-0.4}$ & 0.7$^{+0.3}_{-0.2}$ \\
$\chi^{2}$/dof & 201.7/199 & 273.2/194 \\

 & & \\

{\sc HighEcutPL}: & & \\
$N_{\rm H}$$^{a}$ & 9.7$^{+0.7}_{-0.9}$ & 15.4$^{+2.0}_{-1.7}$ \\
$\Gamma$ & 1.33$^{+0.09}_{-0.16}$ & 0.9$\pm$0.2 \\
$E_{\rm cut}$ (keV) & 12$^{+2}_{-3}$ & 6.0$^{+0.9}_{-0.7}$ \\
$E_{\rm fold}$ (keV) & 21$^{+4}_{-3}$ & 17$^{+5}_{-3}$ \\
$K_{\rm PL}$ ($\times$10$^{-2})$ & 3.4$^{+0.7}_{-0.6}$ & 
	0.43$^{+0.25}_{-0.15}$ \\
$\chi^{2}$/dof & 199.0/198 & 262.7/193 \\

 & & \\

{\sc CompTT}: & & \\
$N_{\rm H}$$^{a}$ & 5.5$^{+0.9}_{-0.7}$ & 11.7$^{+1.7}_{-1.5}$ \\
$T_{\rm 0}$ (keV) & 1.31$^{+0.12}_{-0.14}$ & 1.51$\pm$0.15 \\
$T_{\rm e^-}$ (keV) & 7.8$^{+1.5}_{-1.1}$ & 9.7$^{+2.4}_{-1.6}$ \\
$\tau$ & 4.1$^{+0.7}_{-0.6}$ & 3.5$\pm$0.6 \\
$K_{\rm CompTT}$ ($\times$10$^{-3}$) & 6.1$^{+1.2}_{-1.1}$ & 1.4$\pm$0.3 \\
$\chi^{2}$/dof & 212.7/198 & 261.9/192 \\

 & & \\

{\sc CompST}: & & \\
$N_{\rm H}$$^{a}$ & 10.7$\pm$0.7 & 20.9$\pm$1.5 \\
$T_{\rm 0}$ (keV) & 6.1$^{+0.6}_{-0.5}$ & 7.5$^{+1.3}_{-1.0}$ \\
$\tau$ & 10.4$\pm$0.9 & 8.8$^{+1.3}_{-1.1}$ \\
$K_{\rm CompST}$ ($\times$10$^{-2}$) & 5.2$^{+0.8}_{-0.7}$ & 
	1.9$^{+0.4}_{-0.3}$ \\
$\chi^{2}$/dof & 229.9/199 & 290.0/194 \\

\noalign{\smallskip}
\hline
\noalign{\smallskip}
\hline
\noalign{\smallskip}
$L_{\rm 2-10~keV}$$^{b}$    & 13.5 & 3.7 \\
$L_{\rm 10-100~keV}$$^{b}$  & 24.1 & 7.9 \\

\noalign{\smallskip}
\hline
\hline
\noalign{\smallskip}
\multicolumn{3}{l}{$^{a}$ In units of 10$^{22}$ cm$^{-2}$} \\
\multicolumn{3}{l}{$^{b}$ In units of 10$^{35}$ erg s$^{-1}$} \\
\end{tabular}
\end{center}
\end{table}

\begin{figure*}[t!]
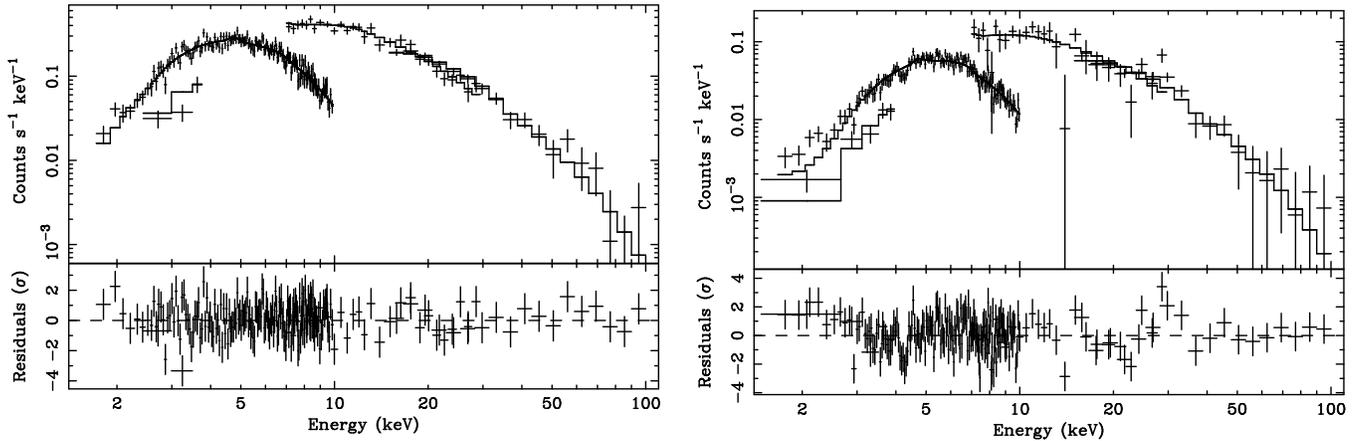

\psfig{figure=aa3654f4l.ps,width=8.7cm,angle=270}
\vspace{-5.7cm}
\hspace{9cm}
\psfig{figure=aa3654f4r.ps,width=8.7cm,angle=270}
\caption[]{{\it Left panel}: 1.5--100 keV X--ray spectrum of 4U 0114+65
obtained with the {\it BeppoSAX} NFIs during the flare reported in Fig. 1
and fitted with a photoelectrically absorbed PL modified with a high
energy exponential cutoff ({\sc HighECutPL} model of Table 2). The
best-fit model is shown as a continuous line. {\it (Right panel)}:
1.5--100 X--ray spectrum of 4U 0114+65 obtained with the {\it BeppoSAX}
NFIs during the low-level state following the flare reported in Fig. 1.
The best-fit model is the same as in the left panel. Note the 
hint of a soft X--ray excess below 3 keV. For both panels, the residuals 
between data and model are shown in units of $\sigma$, and the 
differences in counts s$^{-1}$ keV$^{-1}$ among the various instruments 
are due to different effective detector areas.}
\end{figure*}

To perform the spectral analysis, the NFI pulse-height spectra were
rebinned to oversample by a factor of 3 the full width at half maximum 
(FWHM) of the energy resolution and to have a minimum of 20 counts per bin, 
such that $\chi^2$ statistics could reliably be used. Data were then 
selected, for each NFI, in the energy intervals in which sufficient counts 
were detected from the source and for which the instrument response 
function was well determined. This led us to consider the spectral interval 
1.5--4 keV for the LECS, 1.6--10 keV for the MECS, 7--30 keV for the HPGSPC 
and 15--100 keV for the PDS. We then used the package {\sc xspec} (Dorman 
\& Arnaud 2001) v11.3.1 to fit the resulting broad band energy spectrum.

We included in all fits described here an interstellar photoelectric
absorption column, modeled using the Wisconsin cross sections as
implemented in {\sc xspec} (Morrison \& McCammon 1983) and with solar
abundances as given by Anders \& Grevesse (1989).

When performing the spectral fits, normalization factors were applied to
LECS, HPGSPC and PDS spectra following the cross-calibration tests between
these instruments and the MECS (Fiore et al. 1999). These factors were
constrained to be within the allowed ranges during the spectral fitting,
with the possible exception of the HPGSPC one, with results slightly
($\sim$10\%) higher than the expected range. This shift is however
justified by the fact that the MECS and PDS data, in the ranges in which
they overlap those of the HPGSPC, follow the same bin-to-bin variation
pattern. We can moreover exclude the presence of a contaminating field
source because (i) no catalogued sources with 2--10 keV flux $>$10$^{-13}$
erg cm$^{-2}$ s$^{-1}$, i.e. 10$^{3}$ times fainter than 4U 0114+65, are
present in the HPGSPC field of view (1$^{\circ}$$\times$1$^{\circ}$ FWHM),
(ii) the MECS and PDS instruments do not need such a higher shift 
for their intercalibration factor and, most importantly, (iii) no source 
other than 4U 0114+65 is detected in the field imaged by the MECS (which 
has a size comparable to that of the HPGSPC).

To explore possible spectral differences between the flaring and
the low-level emission from 4U 0114+65 which could be overlooked by the
simple 2--10 keV hardness-intensity ratio analysis of Fig. 2, we divided
the observation into two parts using the data before and after 01:00 UT of
January 27, 1999 (i.e., the 25$^{\rm th}$ hour in the plot in Fig. 1). 
This allowed us to construct broad (1.5--100 keV) `flare' and `low-level' 
post-flare spectra for this source (reported in Fig. 3), therefore making 
the analysis of possible spectral changes described in Sect. 3.1 more 
complete. For these two time-resolved spectra Table 2 reports the models 
along with the corresponding parameters that best fit the data.

In the flaring spectrum, the `classical' phenomenological model of a
power law (PL) plus a high-energy exponential cutoff ({\sc HighECutPL};  
White et al. 1983) provides an acceptable fit to the data, with best-fit
parameters which are similar to, but more accurately determined than
those found e.g. by Bonning \& Falanga (2005) on an {\it INTEGRAL} X--ray
spectrum with similar broadband coverage. An equivalently good fit (see
Table 2) was also achieved using a PL with a simple cutoff ({\sc
CutoffPL}), in the form $N(E) \propto E^{-\Gamma} \cdot e^{-E/E_{\rm cut}}$.

Turning to a more physical description, as suggested by recent
observational (e.g., Masetti et al. 2004; Torrej\'on et al. 2004) and
theoretical (Becker \& Wolff 2005) results, we tried a Comptonization
description of the source flare spectrum via the model by Titarchuk (1994;
{\sc compTT} in {\sc xspec}) and found that it also provided an acceptable
description, with best-fit parameters reported in Table 2. We found no
statistically significant difference in the fit when using different
geometries for the Comptonizing cloud. We therefore decided to assume the
(default) spherical geometry. Also, using the simpler, non-relativistic
modelization by Sunyaev \& Titarchuk (1980; {\sc compST} in {\sc xspec})
for the Comptonization, we obtained acceptable but nevertheless worse
fits (see Table 2); in particular, this description systematically 
underestimates the data of the flare spectrum above 40 keV.

All of these models also describe the low-level spectrum of 4U 0114+65
above 3 keV acceptably well. However, as one can see in Fig. 4, below this
energy an apparent soft X--ray excess is found despite the high value of
$N_{\rm H}$, with the overall 1.5--100 keV fit providing a $\chi^2$/dof
$\sim$ 1.3--1.4 (where `dof' stands for `degrees-of-freedom')  
independently of the chosen model among those listed in Table 2. We tried
to fit this excess with a thermal component, and found that the fit is
marginally improved (to a reduced $\chi^2 \sim$ 1.2, which corresponds to
a 90\% confidence level according to the $F$-test; e.g., Press
et al. 1992) using either a blackbody (BB) with $kT_{\rm th} \sim$ 0.3 keV
and radius $R_{\rm th} \sim$ 200 km, or with a disk-BB model (DBB; Mitsuda
et al. 1984) with similar temperature and radius, or with a hot diffuse
thermal plasma ({\sc mekal} model in {\sc xspec}; Mewe et al. 1985), again
with a similar temperature as of the BB model. In the case in which this
thermal component is added to the Comptonization, an acceptable fit
($\chi^2$/dof = 237.6/191) is also obtained if one forces the thermal
emission to be the source of the Comptonization seed photons by imposing
$kT_{\rm th} \equiv kT_{\rm 0}$. The value of this parameter is again
$\sim$0.3 keV, whereas those of the Comptonization do not differ from
those reported in Table 2.

In Table 2 the unabsorbed luminosities of the source in the two
brightness states and in the 2--10 and 10--100 keV ranges are also
reported, assuming a distance to 4U 0114+65 of 7.0 kpc (Reig et al.
1996). To compute them, the best-fit PL plus high energy exponential
cutoff (i.e., {\sc HighECutPL}) modeling described above as the reference
spectral modelization of the source was used. No substantial differences
in these luminosity values are found if one uses the PL plus simple
cutoff or the Comptonization best-fit models.

Finally, we searched for the presence of an iron emission line around 6.4 
keV (modeled as a Gaussian) and of a CRF at $\sim$22 keV (using the {\sc 
cyclabs} multiplicative model in {\sc xspec}; Mihara et al. 1990;  
Makishima et al. 1990). From our best-fit descriptions there appears to be 
no need to include either of the two spectral features mentioned 
above in the 4U 0114+65 flaring state. Instead, concerning the low-level 
state, while we again did not find any evidence for a CRF, a broad (FWHM 
$\sim$ 1 keV) Fe emission is required with a probability of improvement by 
chance of 5.8$\times$10$^{-5}$.

For each of the two parts of the {\it BeppoSAX} observation we thus
computed the value (when applicable) or the 90\% confidence level upper
limits to the equivalent width (EW) of a Fe emission at 6.4 keV (assuming
different line FWHMs) and to the depth of a CRF at 22 keV (assuming a line
width of 10 keV, as found by Bonning \& Falanga (2005) and which is
typical of magnetic NSs in HMXBs; e.g. Orlandini \& Dal Fiume 2001). 
The
results, reported in Table 3, are independent of the chosen best-fit 
spectral description.

However, from a visual inspection of the right panel of Fig. 4, there 
is a hint of a feature around 20 keV in the low-level spectrum of 4U 
0114+65. In order to check whether
the feature is real or due to a non-perfect modeling of the continuum we
performed a Crab ratio of the PDS spectra for both brightness states. This
technique was used to successfully pinpoint CRFs in numerous X--ray binary
pulsars (see, e.g., Orlandini et al. 1998), and consists of obtaining the
ratio between the count rate spectrum of the source and that of the
featureless power law spectrum of the Crab. This ratio has the advantage
of minimizing the effects due to detector response and calibration
uncertainties. Any real feature present in the source spectrum can then be
enhanced by multiplying the ratio by the functional form of the Crab
spectrum (a power law with photon index $\Gamma$ = 2.1) and by dividing it
by the functional form of the continuum adopted to fit the broad-band
source spectrum, obtaining the so-called Normalized Crab Ratio (NCR; see
Orlandini et al. 1998 for details).  
The NCR performed on both the flare and low-level 4U 0114+65 spectra does
not show any incorrect modeling of the continuum and further excludes the
presence of any absorption feature at $\sim$22 keV.

Also, if we force our low-level spectral data to be described by a 
spectrum like that of the best fit by Bonning \& Falanga (2005), we obtain 
a larger reduced $\chi^2$ ($\sim$1.8), a larger HPGSPC intercalibration 
constant, and a wide fictituous emission bump between 20 and 50 keV, i.e., 
a trend in the fit residuals which is opposite with respect to that found 
by those authors. Because of this, we regard the description of our 
{\it BeppoSAX} low-level data using the best-fit results by Bonning \& 
Falanga (2005) as statistically not satisfactory.

\begin{table}
\caption[]{Values and 90\% confidence level upper limits to the 
EW (in eV) of the 6.4 keV Fe emission line, assuming three different 
fixed line FWHMs, and to the CRF depth assuming an energy $E_{\rm CRF}$ = 
22 keV and a width of 10 keV.}
\begin{center}
\begin{tabular}{l|rrr|c}
\noalign{\smallskip}
\hline
\noalign{\smallskip}
\multicolumn{1}{c|}{Observation} & \multicolumn{3}{c}{Fe line EW (eV)} &
\multicolumn{1}{|c}{CRF depth} \\
\noalign{\smallskip}
\cline{2-4}
\noalign{\smallskip}
 & \multicolumn{1}{|c}{0.1 keV} & \multicolumn{1}{c}{0.5 keV} &
\multicolumn{1}{c|}{1 keV} & \\
\noalign{\smallskip}
\hline
\noalign{\smallskip}

Flare          & $<$61 & $<$93 & $<$141 & $<$0.3 \\
Low state      & 80$\pm$50 & 250$^{+100}_{-90}$ & 580$\pm$200 & $<$0.3 \\

\noalign{\smallskip}
\hline
\noalign{\smallskip}
\end{tabular}
\end{center}
\end{table}

\section{Discussion}

The {\it BeppoSAX} observation of 4U 0114+65 presented in this paper
allowed us to explore the X--ray behaviour of this source in the widest
band thus far achieved (1.5--100 keV). Analysis of the X--ray light curves 
of the object shows that it underwent a large flare, with total duration 
$\sim$4 hours, at the beginning of the pointing. This was followed by a
progressive return of the source to a low-level flux at an intensity about
20 times fainter than the peak of the flare. One can indeed see from Table
2 that the average flux level during the flare is more than $\sim$3 times
that of the average post-flare emission, going from 3.8$\times$10$^{36}$
erg s$^{-1}$ down to 1.2$\times$10$^{36}$ erg s$^{-1}$ in the 2--100 keV
band. Using the orbital ephemeris determined optically by Crampton 
et al. (1985) we found that this flare occurred just after periastron 
passage, at phase $\phi \sim$ 0.05.

The modulation at $\sim$2.7 hours, attributed to the accreting NS spin 
periodicity, is not significantly detected in our data, although 
fluctuations with timescales of $\sim$3 hours could be seen in the 2--10 
keV light curve, and the PSD of the source in this X--ray range indicates 
a break around the corresponding temporal frequency ($\approx$10$^{-4}$ 
Hz). This may be due to the relatively short duration of the observation 
($\sim$3 pulse cycles if one considers the on-source time), to the sparse 
light curve sampling and to the presence of the large flare at the 
beginning of the pointing. X--ray flickering down to timescales of 
minutes was instead found, with its PSD being proportional to 1/$f$, where 
$f$ is the temporal frequency. These short-term variations in the X--ray 
flux may be due to the instability of the accretion process, or to the 
inhomogeneities in the accreting stellar wind (or both; see, e.g., Kaper 
et al. 1993). Thus, both of these long-term (flares) and short-term 
(flickering) `shot-noise' variabilities point to an explanation for this 
X--ray activity as due to random inhomogeneities in the accretion flow 
onto a compact object (e.g. van der Klis 1995).

The possible presence of spectral evolution was investigated through a
2--10 keV hardness-intensity diagram and by accumulating the 1.5--100 keV
X--ray spectra of the flare and of the low-level post-flare emission. The
hardness-intensity diagram does not decisively show appreciable color
evolution in the 2--5 keV and 5--10 keV bands flux; on the other hand, the
comparison of the flare and low-level spectra of 4U 0114+65 as observed by
{\it BeppoSAX} shows that (leaving aside the obvious variation in the
model normalization due to the change in the flux level)  no substantial
modifications in the spectral parameters are apparent, with the notable
exception of the neutral hydrogen column density $N_{\rm H}$.

Indeed, its value nearly doubles at lower emissions independently of
the chosen best-fit spectral modelization. This trend was already
suggested by the {\it EXOSAT}, {\it Ginga} and {\it RXTE} data (Apparao
et al. 1991; Yamauchi et al. 1990; Hall et al. 2000). However, when
compared to those previous observations, we find a larger value for
$N_{\rm H}$. This difference can be explained by the fact that the
$N_{\rm H}$ measurement obtained with {\it BeppoSAX} is substantially
more accurate thanks to the better sensitivity and spectral coverage at
low energies afforded by the {\it BeppoSAX} LECS and MECS.
Thus, in our opinion the $N_{\rm H}$ value obtained with {\it BeppoSAX} 
should be considered as the correct one.

We also note that the $N_{\rm H}$ values we found in both flare and
low-level spectra are substantially higher than the Galactic one along the
4U 0114+65 line of sight, which is $N_{\rm H}^{\rm G}$ =
0.76$\times$10$^{22}$ cm$^{-2}$ according to Dickey \& Lockman (1990).
This fact is also confirmed by the optical $V$-band absorption estimate
along the LSI +65$^{\circ}$010 line of sight made by Reig et al. (1996):  
the value they obtain, $A_V$ = 3.84, corresponds to a $N_{\rm H}$ =
0.69$\times$10$^{22}$ cm$^{-2}$ using the empirical formula by Predehl \&
Schmitt (1995). This is consistent with the Galactic $N_{\rm H}$ value and
substantially smaller than the $N_{\rm H}$ we found from X--ray spectral
fits. This difference is very likely due to the fact that the stellar 
wind absorbs soft X--rays, but does not cause an increase in $E(B-V)$, and 
thus $A_V$, because the wind does not contain dust (see, e.g., Cox et al. 
2005). This fact further suggests the presence of a noticeable and 
variable amount of neutral gas local to the system and very close to the 
NS (possibly, the accretion stream itself).

A variation in the characteristic energy $E_{\rm cut}$ and (to a lesser
extent) in the values of $E_{\rm fold}$ and $\Gamma$ of the {\sc
HighECutPL} model is also found, with these parameters decreasing in
value as the total flux decreases. This trend is opposite to what was 
found by Hall et al. (2000) in the {\it RXTE} data. This may at least 
partly be due to the wider spectral coverage afforded by {\it BeppoSAX}, 
which allowed a better sampling of the X--ray spectral states of 
4U 0114+65.

Concerning the possible presence of a further
component in the softer part of the spectrum (below 3 keV), we note that
Hickox et al. (2004) found that a soft excess, usually modeled using a BB
with $kT_{\rm BB} \sim$ 0.1 keV, is ubiquitous in X--ray pulsars (XRPs)  
and, in particular, in HMXBs hosting pulsating NSs. According to them,
XRPs with luminosities larger than $\sim$10$^{38}$ erg s$^{-1}$ are
produced by reprocessing of hard X--rays emitted by the NS by optically
thick accreting material (e.g., an accretion disk), whereas in cases of
XRPs with luminosities below $\sim$10$^{36}$ erg s$^{-1}$ this mechanism
can be excluded and, most likely, the soft excess arises from different
processes, namely emission from either a photoionized, or collisionally
heated, diffuse gas or from the NS surface. 

Given the size, exceeding the typical NS radius of a factor $\sim$20, 
that we found for the thermal component responsible for the possible soft 
excess in 4U 0114+65, we suggest that a cloud of diffuse plasma around 
the NS produces this excess. 

The Fe emission, its variability and the fact that it is better seen 
during low emission levels were noted by Yamauchi et al. (1990) and 
Hall et al. (2000). This emission line may arise from the heated gas 
cloud depicted above as the possible cause of the soft X--ray excess and, 
as with the continuum emission from this cloud, is apparently better 
detected when 4U 0114+65 is in the low state.

We did not find a CRF at 22 keV in any of the two brightness states; this 
is at variance with the findings of Bonning \& Falanga (2005), who do not 
report any value for the line depth of their CRF detection. Therefore we 
are not able to confirm the estimate of the NS magnetic field strength 
made by these authors and, in turn, the magnetar origin proposed by Li \& 
van den Heuvel (1999) for the NS hosted in this system.

From visual inspection of the right panel of Fig. 4, there is a 
suggestion of a feature at $\sim$20 keV in the low-level spectrum of 4U 
0114+65. Using the NCR technique, we demonstrated in Sect. 3.3 that this 
feature is not due to an incorrect modeling of the X--ray spectral 
continuum, and thus we exclude the presence of any absorption feature at 
$\sim$22 keV in our {\it BeppoSAX} data at both flux levels.

This same situation is observed for another X--ray binary pulsar: Vela 
X--1. As pointed out by Orlandini \& Dal Fiume (2001)  and Coburn et al. 
(2002), in the 15--30 keV energy interval there seems to occur a change of 
slope that is not well described in terms of actual spectral models, 
and that can give rise to residuals that can be misinterpreted as 
absorption features (see also the discussion on the X--ray binary pulsar 
OAO 1657$-$415 in Orlandini et al. 1999).

The above findings suggest the following scenario for 4U 0114+65: the 
accretion stream from the secondary star, in the form of a low angular 
momentum stellar wind as suggested by, e.g., Reig et al. (1996), is 
captured by the accreting NS and creates a plasma cloud around the 
accretor. This cloud thermally emits at a temperature $kT \sim$ 0.3 keV 
and also produces an iron emission line. The thermal emission of the 
cloud is best seen at low hard X--ray flux levels, when this plasma cloud 
is denser, otherwise it is overwhelmed by the more intense underlying 
harder X--ray emission from the NS surface and/or from the NS-cloud 
boundary layer.

The total luminosity of the source is then driven by the density
of the accreting cloud: when higher, it has the double effect of masking 
the inner harder emission and of producing a more intense softer thermal 
emission from its outer parts. Within this scenario, the presence of the 
superorbital period reported in X--rays by Farrell et al. (2005) cannot 
be explained by the presence of a precessing warped accretion disk;
rather, it might be produced by the NS spin axis precession.

\begin{acknowledgements}

This work has made use of the ASI Science Data Centre Archive at
ESA/ESRIN, Frascati, of the NASA's Astrophysics Data System and of the
SIMBAD database, operated at CDS, Strasbourg, France. {\it BeppoSAX} was a
joint program of Italian (ASI) and Dutch (NIVR) space agencies. This
research has been partially supported by ASI. We thank Antonino La Barbera 
for an independent check of the HPGSPC spectra presented in this paper, 
and the referee, Lex Kaper, for several useful comments which helped us to 
improve the paper.

\end{acknowledgements}

\end{document}